\documentclass[preprint,3p,times]{elsarticle}

\usepackage{amsmath,amssymb,amsfonts}
\usepackage{algorithmic}
\usepackage{graphicx}
\usepackage{algorithm,algorithmic}
\usepackage[utf8]{inputenc}
\usepackage{multirow} 
\usepackage{rotating}
\usepackage{subcaption}
\usepackage{tabularx}
\usepackage{booktabs}
\usepackage[pdfencoding=auto, psdextra]{hyperref}

\usepackage{textcomp}

\journal{Nuclear Physics B}

\begin{document}

\begin{frontmatter}

\title{An Explainable Federated Framework for Zero Trust Micro-Segmentation in IIoT Networks} 

\author[COE]{Muhammad Liman Gambo\texorpdfstring{\corref{cor1}}{}} 
\ead{g202320790@kfupm.edu.sa}

\author[COE,ISS]{Ahmad Almulhem}
\ead{ahmadsm@kfupm.edu.sa}

\cortext[cor1]{Corresponding author}

\affiliation[COE]{organization={Department of Computer Engineering, King Fahd University of Petroleum \& Minerals},
            city={Dhahran},
            postcode={31261}, 
            country={Saudi Arabia}}

\affiliation[ISS]{organization={
Interdisciplinary Research Center for Intelligent Secure Systems, King Fahd University of Petroleum \& Minerals},
            city={Dhahran},
            postcode={31261}, 
            country={Saudi Arabia}}

\begin{abstract}

Micro-segmentation as a core requirement of zero trust architecture (ZTA) divides networks into small security zones, called micro-segments, thereby minimizing impact of security breaches and restricting lateral movement of attackers. Existing approaches for Industrial Internet of Things (IIoT) networks often remain centralized, static, or difficult to interpret. These limitations are especially problematic in IIoT environments, where devices are heterogeneous, communication behavior evolves over time, and raw data sharing across sites is often undesirable. Accordingly, we propose EFAH-ZTM, an Explainable Federated Autoencoder-Hypergraph framework for Zero Trust micro-segmentation in IIoT networks. The framework includes a trained federated deep non-symmetric autoencoder that learns behavioral embeddings from distributed clients. After this, kNN-based and Manifold-based hypergraphs capture higher-order relationships among device-flow instances. To generate micro-segments, MiniBatch KMeans and HDBSCAN clustering techniques are applied on the spectral embeddings, while an operational risk score that combines reconstruction error and structural outlierness drives allow/block policy decisions. The trustworthiness of the policy decision is improved through feature-level explanations using LIME and SHAP explainable AI techniques. Experiments on the WUSTL-IIoT-2021 dataset show that HDBSCAN achieved the strongest structural quality, while the manifold-based hypergraph produces the best oracle-aligned security efficacy that reaches a purity of 0.9990 with near-zero contamination. Similarly, the explainability module also showed high fidelity and stability, with surrogate classifier having an accuracy of 0.9927 and stable LIME/SHAP explanations across runs. Moreover, an ablation analysis shows that the federated learning preserves competitive segmentation quality relative to centralized training, and the hypergraph modeling significantly improves structural separation and risk stratification.

\end{abstract}

\begin{keyword}
Explainable AI, Federated Learning, Hypergraph, Industrial IoT, IIoT, Micro-segmentation, Network Security, Zero Trust Architecture

\end{keyword}

\end{frontmatter}


\section{Introduction}
\label{sec:introduction}


Advances in smart devices and wireless communication technologies have led to the increasing adoption of the Internet of Things (IoT) \cite{9509294}. Through sensing and networking capabilities, physical devices are interconnected to each other and to the internet, enabling the collection, sharing and processing of a large amount of data generated by these devices \cite{10601684}, \cite{10440434}, \cite{s22197433}. The applications of IoT have cut across various sectors, including industrial domains such as oil and gas, transportation, power systems, agriculture, etc., formally referred to as Industrial IoT (IIoT) \cite{10122774}. IIoT enables industries to improve their performance, achieve cost reduction, enhance accessibility and scalability, and save time through intelligent automation and analytics\cite{9802088}, \cite{AHMED2023108847}. The market value for IIoT is projected to reach more than USD 100 billion by 2026 \cite{10122774}. 


In the IIoT network, interconnected devices can communicate with each other, sharing and accessing resources from across the network. Without security controls, this presents significant security concerns such as the lateral movement of attackers and privacy issues. This means that when one device is compromised, it can give access to other devices in the flat network \cite{MurshedulArifeen}. To mitigate these threats, the networks are divided into small zones called micro-segments. Micro-segmentation has been identified as one of the core requirements of Zero Trust Architecture (ZTA), a security strategy that assumes no trust and requires that every access request be verified before it is granted  \cite{gambo2025zero}, \cite{LiDi}. The annual revenue for micro-segmentation is estimated to be billions of dollars with a growth rate of more than 20\% \cite{mani2025securing}. 


The security importance of micro-segmentation is generally recognized in  enhancing network security \cite{AndongChen}. It has the ability to reduce the attack surface of networks \cite{MurshedulArifeen}, improve the privacy of sensitive data and systems \cite{HusseinAlshorman}, and provide visibility and control to the increasing amount of east-west traffic \cite{BastaNardine}. While existing micro-segmentation techniques can be effective in a controlled environment, they may not be suitable for IIoT settings which are characterized by heterogeneous devices, limited visibility, and continuously evolving communication behaviors. For example, Software Defined Network-based (SDN) and Virtual Local Area Network-based (VLAN) methods enforce isolation but provide limited behavioral insight and remain largely static \cite{253364}, \cite{LiDi}. Similarly, Machine learning-based (ML) solutions improve automation but lack explainability and adaptability across distributed industrial sites \cite{MurshedulArifeen}, \cite{10864440}. Also, Cloud-native approaches, including intent-driven \cite{ZixuanMa} and configuration-driven \cite{AndongChen}, \cite{mani2025securing} segmentation, assume capabilities unavailable in resource-constrained IIoT systems. These limitations have highlighted the need for a micro-segmentation framework that is suitable for the IIoT environment, trustworthy, and adaptive. Our work addresses this gap through a federated, explainable, and hypergraph-driven ZTA micro-segmentation framework.

Our proposed ZTA micro-segmentation framework builds on the work proposed by Selciya \textit{et al.} \cite{10864440} and Arifeen \textit{et al.} \cite{MurshedulArifeen}. Hence, the key contributions are as follows:

\begin{enumerate}
    \item We propose EFAH-ZTM, a dynamic micro-segmentation framework that autonomously generates security policies for effective micro-segmentation of heterogeneous IIoT devices. This involves systematic integration of existing techniques into end-to-end framework for trustworthy Zero Trust micro-segmentation in IIoT networks.
    \item We implement FedAvg for a Deep Non-Symmetric Auto-Encoder (DNAE) and evaluate under  Dirichlet non‑IID client splits. The DNAE learns device-level behaviors without centralizing sensitive industrial data. This aligns with the distributed nature of the IIoT environment, ensures raw data privacy, and eliminates cross-site data sharing which are critical for industries such as healthcare and manufacturing where the security and privacy of health records and operational process data are critical.
    \item We design an explainable hypergraph clustering method that captures higher-order IIoT relationships and provides interpretable micro-segmentation policies using two eXplainable Artificial Intelligence (XAI) techniques, Local Interpretable Model-agnostic Explanations (LIME) and SHapley Additive exPlanations (SHAP).
    \item We propose a policy generation mechanism based on an operational risk score that enforces allow/deny decisions between micro-segments accordingly. Supplementing the policies with XAI-derived justifications enables security teams to understand why a device or flow is allowed or blocked.
    \item We compare the effectiveness of the proposed EFAH-ZTM framework to ablations (centralized - no FL, no hypergraph) using the WUSTL-IIoT-2021 dataset.
\end{enumerate}


The remainder of the paper is organized as follows. We present a summary of the existing work related to the study in Section~\ref{sec:relatedWork}. Section~\ref{sec:background} presents preliminaries including background on ZTA and micro-segmentation, hypergraph, interpretable segmentation and the threat model considered in the work. We discuss the methodological procedure followed in the study in Section~\ref{sec:methodology}. We discuss the experimental setup including the environmental description, dataset and pre-processing, FL setup, ablations, and performance evaluation in Section~\ref{sec:expSetup}. The results from the study with discussion are presented in Section~\ref{sec:results}. Section~\ref{sec:conclusion} concludes the paper and presents future research.

\section{Related Work}
\label{sec:relatedWork}

Isolating networks into small security zones, through micro-segmentation, represents an important security control with strong security capabilities, including preventing lateral movement of attackers and minimizing attack surface. A number of studies have proposed various techniques for implementing micro-segmentation for conventional networks, cloud environment, and IoT/IIoT networks. In \cite{253364}, Osman \textit{et al.} proposed an SDN-enabled micro-segmentation framework for smart homes. The framework uses device inventory and OpenFlow rules to isolate IoT devices. While effective for small networks, their solution requires centralized edge-cloud components, and policies remain coarse/static compared to workload-level behaviors. Similarly, ML-based micro-segmentation for IIoT is proposed in \cite{MurshedulArifeen} and \cite{10864440}. Arifeen \textit{et al.} \cite{MurshedulArifeen} used OPTICS clustering and decision trees (DT) to automatically segment IIoT traffic and block malicious flows. Whereas Selciya \textit{et al.} \cite{10864440} generate hypergraph clusters using k-NN and deep autoencoders to capture complex IIoT relationships. Although both approaches are promising, the proposed frameworks are centralized contrary to the distributed nature of IIoT network, and are non-explainable, meaning that micro-segmentation decisions cannot be justified or audited. Moreover, a single hypergraph type was used in \cite{10864440}.

A VLAN/VxLAN-based approach is proposed by Li \textit{et al.} \cite{LiDi} which enforces micro-segmentation in cloud data centers. While simple and hardware-agnostic through using industry-standard protocols, the method cannot learn behavioral patterns, segments remain static which does not support dynamic adaptation, and is unsuitable for resource-constrained IIoT environment. An intent-driven micro-segmentation framework is proposed by Ma \textit{et al.} \cite{ZixuanMa} called IMS. IMS defines computable policy algebra to detect inconsistencies and enforce dynamic label-based micro-segmentation in cloud environments. The authors addressed policy inconsistency but did not target the heterogeneous IIoT environment and lack local distributed learning capability.

AutoSeg is another micro-segmentation framework for cloud environment proposed by Andong \textit{et al.} \cite{AndongChen}. It extracts service dependencies from configuration files to build micro-segmentation policies automatically. While highly auditable, it assumes cloud-native application metadata and cannot be applied to the diverse IIoT devices where configurations are not accessible or standardized. Similarly, Mani \textit{et al.} \cite{mani2025securing} proposed a role-based micro-segmentation framework that determines endpoint roles from communication patterns in public clouds. This provides scalability but requires abundant flow logs and does not support privacy-preserving training or explainability which are both essential in IIoT networks. Overall, it is evident that the existing approaches are either network-centric (VLAN/SDN) or target towards cloud networks. Moreover, by focusing on dimensions that are essential for ZT enforcement in IIoT networks, such as adaptability, interpretability, ZT alignment, and operational trustworthiness, Table~\ref{tab:comparisonAnalysis} demonstrates that the existing solutions address only subsets of the requirements for ZT micro-segmentation in IIoT networks. Consequently, through the integration of federated behavioral learning, high-order hypergraph modeling, dynamic adaptability, and explainable risk-aware policy generation, the proposed EFAH-ZTM framework provides a more comprehensive and operationally trustworthy solution specific to the realities of IIoT deployments.

\begin{table*}[t]
\centering
\caption{Comparison of EFAH-ZTM with Existing Micro-Segmentation Frameworks}
\label{tab:comparisonAnalysis}
\renewcommand{\arraystretch}{1.8}
\resizebox{\textwidth}{!}{%
\begin{tabular}{lcccccccc}
\hline
Metric &
Arifeen \textit{et al.} \cite{MurshedulArifeen} &
Osman \textit{et al.} \cite{253364} &
Selciya \textit{et al.}\cite{10864440} &
VLAN/VxLAN \cite{LiDi} &
AutoSeg \cite{AndongChen} &
IMS \cite{ZixuanMa} &
Mani \textit{et al.} \cite{mani2025securing} &
EFAH-ZTM (proposed) \\
\hline
Target Environment &
IIoT &
Smart Home IoT &
IIoT &
Cloud DC &
Cloud &
Cloud &
Cloud &
IIoT \\
Learning Paradigm &
Centralized ML &
Rule-based &
Centralized ML &
Rule-based &
Config-based &
Logic-based &
Centralized ML &
Federated ML \\
Privacy Preservation &
$\times$ &
$\times$ &
$\times$ &
$\checkmark$ &
$\checkmark$ &
$\checkmark$ &
$\times$ &
\textbf{$\checkmark$} \\
behavioral Representation Learning &
$\checkmark$ &
$\triangle$ &
$\checkmark$ &
$\times$ &
$\times$ &
$\times$ &
$\checkmark$ &
\textbf{$\checkmark$} \\
High-Order Relationship Capture &
$\times$ &
$\times$ &
$\triangle$ &
$\times$ &
$\times$ &
$\times$ &
$\triangle$ &
\textbf{$\checkmark$} \\
Dynamic Adaptability &
$\triangle$ &
$\checkmark$ &
$\checkmark$ &
$\times$ &
$\triangle$ &
$\checkmark$ &
$\checkmark$ &
\textbf{$\checkmark$} \\
Policy Interpretability &
Low &
Medium &
Low &
Medium &
High &
High &
Medium &
Very High \\
Risk Quantification &
$\times$ &
$\times$ &
$\times$ &
$\times$ &
$\times$ &
$\times$ & 
$\times$ &
$\checkmark$\\
Zero Trust Alignment &
$\triangle$ &
$\triangle$ &
$\checkmark\checkmark$ &
$\triangle$ &
$\checkmark$ &
$\checkmark\checkmark$ &
$\checkmark$ &
\textbf{$\checkmark\checkmark\checkmark$} \\
IIoT Suitability &
$\checkmark$ &
$\triangle$ &
$\checkmark$ &
$\times$ &
$\times$ &
$\times$ &
$\times$ &
\textbf{$\checkmark\checkmark$} \\
Operational Trustworthiness &
Low &
Medium &
Low &
Medium &
High &
High &
Medium &
Very High \\
\hline
\end{tabular}
}
\vspace{0.2cm}

\raggedright
\footnotesize
Legend:
$\checkmark\checkmark\checkmark$ = Strong/Native support,\;
$\checkmark\checkmark$ = Strong support,\;
$\checkmark$ = Supported,\;
$\triangle$ = Partial support,\;
$\times$ = Not supported,\;
DC = Data Center.
\end{table*}

\section{Preliminaries}
\label{sec:background}


\subsection{Zero Trust Architecture and Micro-segmentation}
Traditionally, networks are protected using firewalls, virtual private networks (VPN), and intrusion detection systems (IDS) \cite{AroraSunil}, \cite{NaharNurun}. These security solutions protect the inside of a network from the outside by implicitly assuming that everything within the network is secure and trustworthy. However, security issues such as insider threat, lateral movement of attackers, and the increasing frequency and sophistication of cyber attacks require a robust and adaptive approach to network security \cite{ZillahAdahman}. Hence, ZTA has emerged as a new concept to access control that emphasizes continuous authentication and conditional authorization under the principle of "never trust, always verify" \cite{SyedNaeem}. It has several application requirements including micro-segmentation and various technologies that enable its deployment \cite{gambo2025zero}. Moreover, based on the proposal of the National Institute of Standards and Technology (NIST), ZTA consists of three main components, which are the policy engine (PE), the policy administrator (PA), and the policy enforcement point (PEP) \cite{NIST_ZTA}, \cite{gambo2025zero}. The PE and the PA are collectively called policy decision point (PDP). Also, the global market value of Zero Trust is projected to reach 133 billion US dollars by 2032 \cite{statista2024zerotrust}.

There exist several conventional network segmentation approaches, such as VLANs and Subnets, Firewalls, and Physical segmentation. While they are effective to the level of their implementation, they each have their individual limitations. For example, with VLANs, they divide networks based on physical location (switch port) or function which is easy to implement, however, VLANs policies are static and reconfiguring to meet evolving demand may be time-consuming and complex, and prone to misconfiguration errors. They also do not provide visibility into application-level communication or control traffic between devices residing within the same VLAN. Similarly, Firewalls are essential tools for establishing network security controls and help to examine traffic moving in and out of a network. But deploying them in a distributed IIoT environment can present significant complexity and high cost because it means deploying a large number of firewalls for the different network segments involved. In addition, managing the massive number of firewall rules can become a security and management burden. Physical segmentation, on the other hand, offers the highest level of isolation where the network is separated using dedicated network hardware (e.g., separate switches, routers). However, this approach is expensive and inflexible, and may be impractical for a large and rapidly evolving network environment like IIoT. Therefore, micro-segmentation serves as a dynamic and adaptable technique to bridge the limitations of the existing network segmentation techniques. 

The implementation of micro-segmentation can take one of three broad forms according to Zixuan \textit{et al.} \cite{ZixuanMa}. These include: (1) IP-based micro-segmentation where segmentation rules are defined based on IP addresses of devices; (2) Group-based micro-segmentation where devices with the same access control specifications are clustered into logical groups and rules are defined per group; (3) Label-based micro-segmentation in which devices are assigned labels, and rules are defined and managed based on the presence and value of these assigned labels.

\begin{figure}[b!]
    \centering
    \includegraphics[scale=0.6]{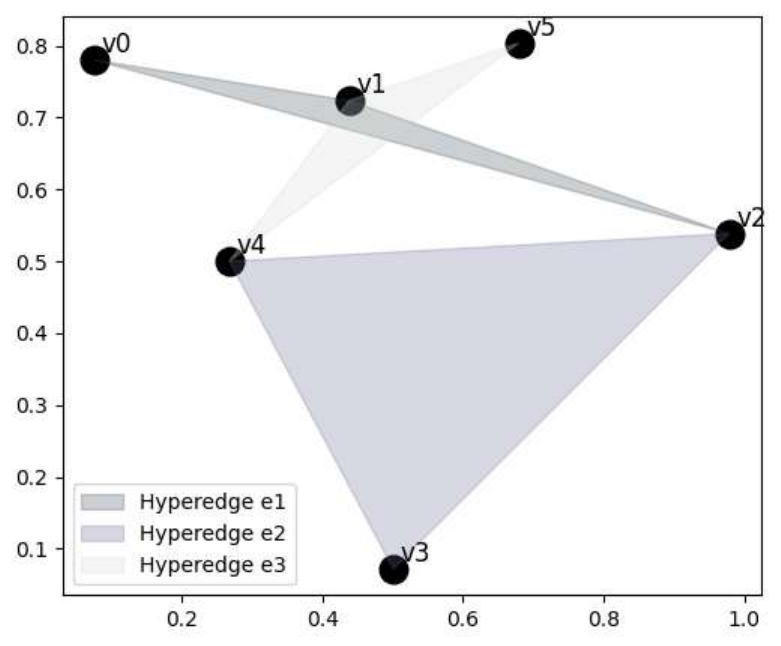}
    \caption{Hypergraph Visualization with Hyperedges}
    \label{fig:hypergraph}
\end{figure}

\subsection{Hypergraph Modeling for IIoT behavioral Structure}

Micro-segmentation in the IIoT environment requires a modeling framework that is capable of capturing the multi-device, multi-protocol, and multi-context interactions which naturally arise in cyber-physical operations \cite{10864440}. Traditional graph models, which encode only pairwise device relations, are fundamentally limited in this aspect of modeling high-order relationships \cite{lotito2024hyperlink}. This is because they are unable to represent simultaneous, higher-order dependencies such as coordinated sensor–controller-actuator loops or shared service–level interactions across multiple devices \cite{niketh2025game}. Modeling of high-order relationships is essential whenever relationships are many-to-many, non-linear, or context-dependent, since converting them into simple graphs either destroys structural semantics or creates artificially dense cliques that distort downstream analysis \cite{9264674}, \cite{9795251}. Hypergraphs overcome this limitation by allowing a single hyperedge to connect an arbitrary number of vertices as shown in Fig.~\ref{fig:hypergraph}, thereby encoding group-level behavioral patterns \cite{10.1145/3605776}, \cite{10864440}.

In the hypergraph formulation adopted in EFAH-ZTM, each IIoT device is represented as a vertex $v \in V$, and hyperedges $e \in E$ capture cohesive behavioral groups such as devices that behave similarly, share protocol usage, or exhibit similar latent representations derived from the federated autoencoder. The kNN-based hypergraph construction method \cite{6570719} used generates hyperedges based on local similarity in latent behavioral space. More specifically, for each device-flow embedding produced by the federated autoencoder, a hyperedge is formed by connecting the point to its \textit{k nearest neighbors}. Manifold-based hypergraph construction method \cite{10903483}, on the other hand, constructs hyperedges through diffusion affinities along the underlying data manifold. Using a heat-kernel similarity and a \textit{t-step} diffusion operator, each device-flow instance forms a hyperedge with the nodes that exhibit the highest diffusion connectivity.

\subsection{Interpretable Segmentation and Explainable AI}

Building on the hypergraph construction which encodes multi-device behavioral relationships, the constructed hypergraph produces a spectral embedding that directly enables effective clustering. Clustering is an unsupervised ML-based method that discovers patterns in a dataset and automatically groups similar data points together \cite{app12052405}. There are various clustering algorithms available, and in this study \textit{MiniBatch KMeans} \cite{10.1145/1772690.1772862} and \textit{Hierarchical Density-Based Spatial Clustering of Applications with Noise (HDBSCAN)} \cite{campello2013density} are used.

\textit{MiniBatch KMeans} is an improved version of the \textit{k-Means} algorithms, and uses small batches to provide centroid-based partitioning while reducing computation and  convergence time especially in large datasets \cite{9702823}. However, as a partition-based method, \textit{MiniBatch KMeans} assumes globular cluster geometry and may struggle when IIoT behaviors form irregular or variable-density patterns \cite{app12052405}. \textit{HDBSCAN} as a density-based clustering algorithm addresses these limitations by identifying clusters of arbitrary shape while distinguishing noise points. It can adapt to heterogeneous device behaviors, distinguishing tightly coupled operational groups from sparse or anomalous behaviors \cite{computation13060144}. Moreover, its ability to assign noise labels is especially valuable in Zero Trust settings, where it enables explicit treatment of uncertain or suspicious samples rather than forcing them into benign clusters.


However, effective Zero Trust micro-segmentation requires more than the ability to form coherent clusters. It also requires the capacity to interpret and justify the assignment of devices or flows to specific micro-segments. This builds trust and ensures effective policy enforcement \cite{11023864}. Integrating an XAI layer in the micro-segmentation pipeline attaches feature-level explanations to every micro-segment assignment and policy rule. Accordingly, two well-known XAI techniques are used: \textit{LIME} \cite{ribeiro2016should} and \textit{SHAP} \cite{lundberg2017unified}. Based on a perturbation mechanism, \textit{LIME} trains a local surrogate model to interpret individual predictions by approximating the decision boundary of the black-box predictor \cite{10440604}. While \textit{SHAP}, driven by the concept of game theory, represents the average contribution of a feature across every subset of features in which it can be included in a prediction \cite{10725120}. In this study, a surrogate classifier, trained on the spectral embeddings, enables these techniques to reveal the most influential features driving segmentation decisions. Building this explainability feature into the micro-segmentation pipeline transforms the policy generation from a black-box process into a transparent, auditable decision system. Here, administrators can gain insight into why a micro-segment is formed, why a device is placed within it, and why an associated communication flow results in an Allow or Block action.

\subsection{Threat Model}

Securing IIoT networks requires a clear definition of the adversarial capabilities, system trust boundaries, and the defensive objectives guiding the micro-segmentation. Here, we present our threat model, which follows the attacker behavior patterns described in MITRE ATT\&CK \cite{mitre_attack_ics_tactics} for industrial control systems (ICS).

\subsubsection{Attacker Capabilities} We assume a network-borne adversary capable of compromising a single IIoT device or flow source. This is consistent with ICS attack patterns such as initial access, command execution, and pivoting. After the attacker has gained initial access, they can:

\begin{itemize}
    \item Inject arbitrary network flows to mimic benign communications.
    \item Attempt lateral movement through scanning for reachable devices or exploiting the implicit trust relationships that exist between devices.
    \item Modify the behavioral characteristics of the compromised device, such as communication frequency or packet-level features.
\end{itemize}

However, despite having these capabilities, the attacker:
\begin{itemize}
    \item Does not possess global visibility of the entire distributed network.
    \item They cannot compromise the federated learning process such as tampering with the global aggregation server or corrupting the training pipeline by performing attacks such as FL poisoning. These adversaries fall outside the scope of this work and are addressed separately in secure FL literature.
    \item They cannot corrupt or compromise the policy enforcement layer once micro-segmentation rules are deployed.
\end{itemize}

\subsubsection{Defender Assumptions} While following ZTA principles and industry segmentation guidelines such as the IEC 62443 Zone-Conduit model \cite{isagca_62443_standards} at the defender level, we assume the following:

\begin{itemize}
    \item The end devices may be compromised, but their local behavioral telemetry remains observable to the federated model.
    \item The federated aggregator is honest-but-curious. This means it will truthfully perform correct model averaging and maintain model integrity.
    \item The training data remain local; safeguarding raw data and ensuring that attackers cannot observe or modify data from other sites.
    \item The hypergraph construction, spectral embedding, and clustering pipeline remain uncompromised; faithfully representing the behavioral structure.
    \item The PE (micro-segmentation enforcement) is trusted to apply generated Allow/Block rules with integrity; consistent with ZTA enforcement-plane requirements.
\end{itemize}

\subsubsection{Security Goals} Through the proposed framework, we aim to enforce zero trust (ZT) micro-segmentation and achieve the following security goals:


\begin{itemize}
    \item Lateral movement containment: One of the key security goals of the framework is to prevent an attacker who has compromised a single device from traversing or moving laterally to other micro-segments. Accordingly, the hypergraph-guided clustering establishes behavioral boundaries, and the micro-segmentation rules ensure that communication is restricted to behaviorally justified, intra-segment devices only.

    \item High-fidelity behavioral micro-segmentation: Another important security goal of the framework is ensuring that benign devices are grouped together while malicious or anomalous ones are isolated. To achieve this, metrics such as cluster purity and contamination are used for offline evaluation to quantify how effectively segments separate benign from malicious behaviors (Section~\ref{subsec:perfEval}).

    \item Explainability for trustworthy enforcement: Through the integration of XAI techniques into the micro-segmentation policy generation pipeline ensures interpretable and trustworthy enforcement. This provides feature-level justifications for every Allow or Block decision generated by the PE.
\end{itemize}

\section{Methodology}
\label{sec:methodology}
\begin{figure}[H]
    \centering
    \includegraphics[width=0.98\linewidth]{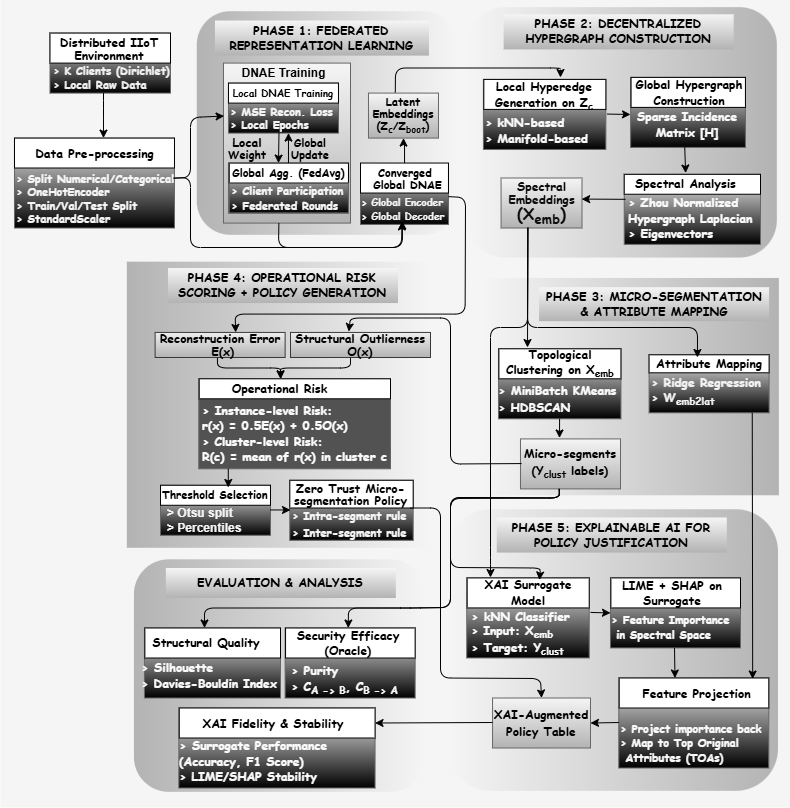}
    \caption{End-to-end Architecture of Proposed EFAH-ZTM Framework}
    \label{fig:EFAHFramework}
\end{figure}

This section describes the multi-stage workflow followed for the proposed EFAH-ZTM framework. Figure~\ref{fig:EFAHFramework} shows the summarized workflow for the methodology. It involves federated representation learning to learn a privacy-preserving, low-dimensional representation of network flow data; decentralized hypergraph construction and spectral embedding to capture the intrinsic, multi-way topology of the data points in the latent space; micro-segmentation to discover micro-segments (clusters); quantifying operational risk score associated with each micro-segment and policy generation; and human-readable interpretation for micro-segmentation policies using explainable AI. These are discussed in detail in the following subsections.

In the experiment, we consider an IIoT environment distributed across $K=10$ sites (clients), for example, plants or zones. Each client $k \in \{1, \dots, K\}$ holds a local dataset:

\begin{equation}
\mathcal{D}_k = \{x_{k,i}\}_{i=1}^{n_k}, \quad x_{k,i} \in \mathbb{R}^n,
\label{eq:datadesc}
\end{equation}

where each $x_{k,i}$ represents a sample generated by an IIoT device and described by $n$ features after preprocessing.

\subsection{Federated Representation Learning}
To align with the distributed nature of the IIoT environment and also prevent the sharing of raw data, we used federated learning (FedAvg) to train a deep non-symmetric autoencoder (DNAE) model for EFAH-ZTM. This means that instead of centralizing the raw traffic logs, each client site trains a local DNAE model on its local data partition. This approach reflects the IIoT setting consisting of dispersed IIoT devices or small but connected industrial divisions within a large industry.

\subsubsection{DNAE Model Architecture}

A DNAE model is developed which comprises of an encoder ($E_{\theta} : \mathbb{R}^{d} \to \mathbb{R}^{p}$) and decoder ($D_{\phi} : \mathbb{R}^{p} \to \mathbb{R}^{d}$) architecture with layer configuration as shown in Table~\ref{tab:dnae_architecture}. Given an input $x$, the latent embedding is $z = E_{\theta}(x)$, the reconstruction is $\hat{x} = D_{\phi}(z)$, and $d$ and $p$ are the number of features in the input and latent space, respectively. In addition, a \textit{LeakyReLU} activation function is used for all hidden layers, with a linear output layer that reconstructs the original feature vector. Also, the model is trained to minimize \textit{Mean Squared Error (MSE)} represented in Eq.~\ref{eq:recError}. This encourages the latent space to capture the intrinsic behavioral structure of a device flow.

\begin{equation}
E(x_i) = \frac{1}{d} \|x_i - D_\phi(E_\theta(x_i))\|^2_2
\label{eq:recError}
\end{equation}

\subsubsection{Federated Training using FedAvg}

In the training process, we  partitioned the dataset into $N=10$ virtual clients, and each client trains the DNAE locally for $r=3$ local epochs. The server aggregates the local model updates from clients using TFF's Weighted FedAvg and the process repeats for $R=50$ federated rounds, aggregating and redistributing the global model update to clients. At the end, the converged Global Encoder ($\mathbf{E}_{global}$) is extracted and used to generate the latent representations ($\mathbf{Z}_{boot}$) which serve as the basis for building the hypergraph that captures IIoT device and flow relationships.

\begin{table}[ht]
    \centering
    \caption{DNAE Model Architecture Specifications}
    \label{tab:dnae_architecture}
    \small
    \begin{tabular}{lllc}
        \toprule
        \textbf{Block} & \textbf{Layer Type} & \textbf{Specifications / Parameters} & \textbf{Output Channels} \\ 
        \midrule
        \multirow{5}{*}{Encoder} & Conv1d + BN & Kernel: 3, Stride: 1, Padding: 1 & 256 \\
                                 & MaxPool1d   & Kernel: 2, Stride: 2            & 256 \\
                                 & Conv1d + BN & Kernel: 3, Stride: 1, Padding: 1 & 128 \\
                                 & MaxPool1d   & Kernel: 2, Stride: 2            & 128 \\
                                 & Conv1d + BN & Kernel: 3, Stride: 1, Padding: 1 & 64 \\
                                 & MaxPool1d   & Kernel: 2, Stride: 2            & 64 \\
                                 & Conv1d + BN & Kernel: 3, Stride: 1, Padding: 1 & 32 \\
                                 & MaxPool1d   & Kernel: 2, Stride: 2            & 32 \\
                                 & Conv1d + BN & Kernel: 3, Stride: 1, Padding: 1 & 16 \\
                                 & MaxPool1d   & Kernel: 2, Stride: 2            & 16 \\
        \midrule
        \multirow{2}{*}{Latent}  & FC (Latent) & In: 16, Out: 25                 & - \\
                                 & FC (Up)     & In: 25, Out: 16                 & - \\
        \midrule
        \multirow{4}{*}{Decoder} & Conv1d + BN & Kernel: 3, Stride: 1, Padding: 1 & 16 \\
                                 & Upsample    & Scale Factor: 2.0 (Nearest)     & 16 \\
                                 & Conv1d + BN & Kernel: 3, Stride: 1, Padding: 1 & 32 \\
                                 & Upsample    & Scale Factor: 2.0 (Nearest)     & 32 \\
                                 & Conv1d + BN & Kernel: 3, Stride: 1, Padding: 1 & 64 \\
                                 & Upsample    & Scale Factor: 2.0 (Nearest)     & 64 \\
                                 & Conv1d + BN & Kernel: 3, Stride: 1, Padding: 1 & 128 \\
                                 & Upsample    & Scale Factor: 2.0 (Nearest)     & 128 \\
        \midrule
        \textbf{Output}          & FC (Out)    & In: 2048, Out: 48               & - \\
        \bottomrule
    \end{tabular}
\end{table}

\subsection{Decentralized Hypergraph Construction and Spectral Embedding}

At this stage, we build decentralized hypergraphs that encode multi-way relationships between samples in the latent space through local summarization and global hypergraph construction. 

\subsubsection{Local Hyperedge Generation}
EFAH-ZTM supports two modes for hyperedge construction. These are kNN-based hypergraphs and Manifold-based hypergraphs. For each mode, hyperedges are constructed locally per client and aggregated into a unified hypergraph Incidence Matrix ($\mathbf{H}$). In kNN-based hypergraphs, each client identifies, within its local latent data ($\mathbf{Z}_c$), the \textit{top-k latent neighbors} for every sample. Accordingly, each sample along with its $k$ neighbors, forms a single hyperedge, and this captures local continuity and connectivity in the latent space. Similarly, manifold-based hypergraphs use diffusion distances computed from the latent space affinities to capture non-linear relationships among samples and define hyperedges.

\subsubsection{Global Hypergraph Representation}
The locally formed hyperedges are aggregated globally to construct the sparse \textit{Incidence Matrix ($\mathbf{H}$)}, where each column corresponds to a hyperedge. Subsequently, using ($\mathbf{H}$) and optional edge weights, we then compute the Zhou \textit{Normalized Hypergraph Laplacian} ($\mathbf{L}$) in Eq.~\ref{eq:laplacian}.

\begin{equation}
\mathbf{L} = \mathbf{I} - \mathbf{D}_v^{-1/2} \mathbf{H} \mathbf{W} \mathbf{D}_e^{-1} \mathbf{H}^T \mathbf{D}_v^{-1/2}
\label{eq:laplacian}
\end{equation}

where $\mathbf{D}_v$ and $\mathbf{D}_e$ are diagonal matrices of node and hyperedge degrees, respectively. This Laplacian forms the foundation for computing hypergraph spectral embeddings.

\subsubsection{Spectral Embedding}
The structural information encoded in $\mathbf{L}$ is transformed into a Euclidean feature vector space through Spectral Embedding. This is critical for enabling subsequent density-based clustering in a space that respects the non-linear, multi-way relationships discovered by the hypergraph. The embedding process is achieved by solving the generalized eigenvalue problem for the Laplacian matrix (\ref{eq:gLaplacian}:

\begin{equation}
\mathbf{L} \mathbf{v} = \lambda \mathbf{D}_v \mathbf{v}
\label{eq:gLaplacian}
\end{equation}

We use the \textit{Locally Optimal Block Preconditioned Conjugate Gradient (LOBPCG)} solver to compute the eigenvectors ($\mathbf{v}$) corresponding to the smallest eigenvalues ($\lambda$). This solver is well suited for large, sparse Laplacian matrices and offers improved numerical stability compared to ARPACK-based methods such as \textit{eigsh}. We discard the trivial first eigenvector (where $\lambda \approx 0$) and select the subsequent $d_{emb}=10$ eigenvectors, which represent the optimal embedding dimensions. These selected eigenvectors form the Spectral Embedding Matrix ($\mathbf{X}_{emb}$), which exposes the topological structure suitable for clustering.

\subsection{Micro-segmentation and Attribute Mapping}
\subsubsection{Micro-segmentation}
This step represents the core of the ZTA micro-segmentation enforcement where the final micro-segments are defined. In the proposed EFAH-ZTM framework, two clustering techniques, \textit{MiniBatch KMeans and HDBSCAN}, were identified and applied to the spectral feature space ($\mathbf{X}_{emb}$) to form the dynamic micro-segments. MiniBatch KMeans is a fast and scalable clustering technique which is valuable when computational resources are constrained or a fixed number of segments is required. While HDBSCAN is a dense topological clustering technique that has the ability to automatically discover clusters of varying density and isolate outliers as noise (labeled $-1$).

The output from either method, which is a vector of cluster labels ($\mathbf{Y}_{clust}$), defines the final dynamic micro-segments. This serves as the basis for the micro-segmentation policies in the policy generation phase.

\subsubsection{Attribute Mapping}

Maintaining the crucial link between the final clustering results and the initial compressed feature space ($\mathbf{Z}_{boot}$) requires a mapping layer to be explicitly modeled. To achieve this, we trained a \textit{Ridge Regression} model that predicts the latent features ($\mathbf{Z}_{boot}$) from the spectral features ($\mathbf{X}_{emb}$). This results in the linear transformation matrix ($\mathbf{W}_{emb2lat}$) and the bias vector ($\mathbf{b}_{emb2lat}$). In addition, the developed model ensures a stable functional relationship: $\mathbf{Z}_{boot} \approx \mathbf{X}_{emb} \mathbf{W}_{emb2lat} + \mathbf{b}_{emb2lat}$. Moreover, the \textit{Attribution Map} ($\mathbf{W}_{emb2lat}$) is essential, because it allows subsequent feature importance vectors generated in the $\mathbf{X}_{emb}$ space by XAI techniques to be reliably projected back into the $\mathbf{Z}_{boot}$ space, and hence preserves semantic meaning closer to the raw input features.

\subsection{Operational Risk Scoring}
\label{subsec:opRisk}

In this section, we describe the operational risk score we used in the proposed EFAH-ZTM framework, which provides a continuous measure of risk suitable for fine-grained, risk-aware policy enforcement. It aggregates complementary, unsupervised signals that capture behavioral deviation measured by reconstruction error from the federated autoencoder, and structural confidence measured in the micro-segmentation embedding space.

Let $x_i \in \mathbb{R}^d$ denote a preprocessed feature vector corresponding to a network flow instance. The operational risk score assigned to $x_i$ is defined in Eq.~\ref{eq:instanceRS}:

\begin{equation}
r(x_i) = w_1 \tilde{E}(x_i) + w_2 \tilde{O}(x_i)
\label{eq:instanceRS}
\end{equation}

where $w_1, w_2\geq 0$ are weighting coefficients satisfying $\sum_j w_j = 1$ and $\tilde{E}(x_i)$, and $\tilde{O}(x_i)$ are normalized risk components described below. Moreover, equal weights ($w_1 = w_2 = 0.5$) are used to avoid bias toward any single signal.

\begin{itemize}
    \item behavioral Deviation through Reconstruction Error $(w_1 \tilde{E}(x_i))$: This component captures the deviation from the learned normal behavior using the DNAE. Given $x_i$, its reconstruction error is computed as shown in Eq.~\ref{eq:recError}. High reconstruction error indicates that the observed behavior deviates from dominant patterns learned during FL training, suggesting anomalous or previously unseen activity. 

    \item Structural Outlierness from Micro-Segmentation $(w_2 \tilde{O}(x_i))$: This measures topological confidence with respect to the discovered micro-segments and captures how weakly an instance is embedded within its assigned behavioral group. For Mini KMeans-based clustering, outlierness is defined as the Euclidean distance to the corresponding cluster centroid, while for HDBSCAN-based clustering, the native outlier or noise score is used. Instances labeled as noise are treated as maximally outlying.

\end{itemize}

To ensure comparability across heterogeneous scales, each component is robustly normalized using median and interquartile range (IQR) statistics. To obtain a bounded risk component, we clip the robust z-score and apply a sigmoid squashing. At the end, we shift the squashed output so that values at or below the median contribute approximately zero risk, and only above-typical values contribute positively. As a result, the final instance-level operational risk is bounded $r(x_i) \in [0, 1]$. Moreover, for policy enforcement, instance-level risk scores are aggregated at the micro-segment level as described in Eq.~\ref{eq:clusterRisk}:

\begin{equation}
    R(c) = \mathbb{E}_{x_i : c_i = c}[r(x_i)]
\label{eq:clusterRisk}
\end{equation}

Cluster-level risk provides a stable, collective assessment of behavioral trustworthiness and forms the basis for risk-aware intra-segment communication policy.

\subsection{Policy Generation}
\label{subsec:policyGen}
The policy rules are dynamically generated, implementing the ZTA principle of granular, risk-aware micro-segmentation. They are structured using cluster membership and a quantified operational risk score (Section~\ref{subsec:opRisk}). We modeled two types of traffic, those initiated and intended for within a micro-segment communication (intra-segment), and those targeted for another micro-segment (inter-segment).

\subsubsection{Intra-segment Policy (Allow/Deny Logic)} A risk-based approach is used to evaluate traffic flows between devices belonging to the same micro-segment (SRC\_CID = DST\_CID) to determine the appropriate action.
\begin{itemize}
    \item Conditional Allow: Under this policy, intra-segment communication is allowed only when the corresponding micro-segment exhibits low operational risk. More explicitly, because each instance risk $r(x_i)$ is constructed from robustly normalized and sigmoid-squashed components, $r(x_i) \in [0, 1]$, values near 0 correspond to typical behavior, and values closer to 1 indicate increasingly abnormal behavior (for example, due to high reconstruction error and/or strong embedding outlierness). Accordingly, we define a cluster threshold $\tau_c \in [0, 1]$ where clusters with $R(c) \leq \tau_c$ are considered operationally safe and are granted conditional intra-cluster communication, and clusters with $R(c) > \tau_c$ are treated as high-risk and are therefore blocked.

    The threshold $\tau_c$ is selected from the empirical distribution of cluster risk scores $\{R(c)\}$ which are computed on the dataset test split. We generate candidate thresholds using (i) interpretable percentile cutoffs (for example, $p50$--$p95$) over $R(c)$ and (ii) Otsu’s method, which identifies a histogram-based split that maximizes between-group separation of low- versus high-risk micro-segments. Through this approach, it provides an operationally transparent basis for choosing the $\tau_c$ under different security postures.
    
    \item Segment Isolation: In a situation whereby the calculated operational risk score for a micro-segment exceeds the acceptable threshold, the communication is blocked. The policy aims to contain situations with elevated behavioral risk within a micro-segment.
\end{itemize}

\subsubsection{Inter-segment Policy (Default Deny)} Under this policy, communication attempts between devices belonging to different micro-segments  (SRC\_CID $\neq$ DST\_CID)  are strictly blocked. This default deny policy reduces the network's attack surface and minimizes the potential for lateral movement across segregated security zones. Moreover, it is unconditional and independent of the operational risk score, consistent with the “never trust, always verify” principle of ZTA.


However, the segment isolation and inter-segment policies may represent an upper bound and could be modified depending on organizational need and risk tolerance.

\subsection{Explainable AI (XAI) for Policy Justification}
\label{subsec:XAIforPolJust}
The generated micro-segmentation policy rules are improved with human-interpretable justifications using LIME and SHAP. They provide an insight, through feature-level explanations, on why specific flows or devices were placed into particular micro-segments.

\subsubsection{Surrogate Classifier:} 
To explain the micro-segmentation assignments, a kNN model \textit{Surrogate Classifier} is trained using $\mathbf{X}_{emb}$ as input features and $\mathbf{Y}_{clust}$ as target classes. The model approximates the decision boundary of the clustering techniques used and provides the predictable interface for XAI techniques. This implemented component of the proposed EFAH-ZTM framework ensures that the generated micro-segmentation policies are transparent, and trustworthy. After training the classifier, LIME and SHAP are applied to determine the feature importance vectors in the spectral space $\mathbf{X}_{emb}$ that drive a specific micro-segmentation decision.

\subsubsection{Attribute Mapping:}
The generated feature importance must be translated from the spectral domain back to the meaningful attributes in the network domain. First, the importance vectors are projected back into the latent space ($\mathbf{Z}_{boot}$) using the pre-trained $\mathbf{W}_{emb2lat}$ matrix. Then, these influential latent representations are then mapped to the most highly-correlated \textit{Top Original Attributes(TOAs)} (e.g., rate, dur, proto). Finally, these TOAs provide explicit, auditable justification that is appended to every rule in the final micro-segmentation policy. These explanations enable security teams to immediately understand and verify why a specific device or flow is either allowed or blocked.

\begin{table*}[h!]
\centering
\caption{Parameters and Hyperparameters Used in EFAH-ZTM}
\label{tab:parameters}
\renewcommand{\arraystretch}{1.5}
\small 
\begin{tabularx}{\textwidth}{l p{3.5cm} X}
\hline
Category & Parameter / Hyperparameter & Value / Setting \\ 
\hline

General Setup & Number of clients & 10 \\ 
              & Dirichlet concentration parameter ($\alpha$) & 0.7 \\
              & Client Participation & 0.8 \\ 

Pre-processing 
              & Encoding & OneHotEncoder \\
              & Scaling & StandardScaler \\ 

DNAE Architecture 
                  & Latent dimension & 25 \\
                  & Activation functions & LeakyReLU, Linear \\
                  & Loss function & MSE \\

Federated DNAE 
               & Client optimizer & Adam (lr = 1e-3) \\
               & Server optimizer & SGD-Momentum (lr = 1.0) \\
               & Local epochs & 3 \\
               & Batch size & 64 \\
               & Federated rounds & 50 \\ 

Hypergraph Construction & Modes & knn\_only and manifold\_hypergraph \\ 
                        & kNN neighbors & 12 \\

Spectral Embedding & Embedding dimensions & 10 \\ 

Clustering & Clusterer & HDBSCAN and MiniBatch KMeans \\ 
           & HDBSCAN parameters & min\_cluster\_size=auto, min\_samples=None, metric=euclidean \\
           & KMeans clusters & 500 \\ 

Policy Generation 
                  & Intra-cluster rule & Allow if $R(c) \leq \tau_{c}$ \\
                  & Inter-cluster rule & Block \\ 
Surrogate Model & kNN neighbors & 25\\
                    & Metric & euclidean \\
                   & Weights & distance \\
                   \hline

\end{tabularx}
\end{table*}

\section{Experimental Setup}
\label{sec:expSetup}

\subsection{Environmental Description and Packages}

All experiments for the proposed EFAH-ZTM framework were conducted in a \textit{Google Colab environment}. This provides a managed and cloud-based execution platform with access to CPUs/GPUs. We selected Google Colab to ensure ease of reproducibility, and accessibility for other researchers wishing to replicate or extend the proposed experiments. Moreover, we used \textit{Python 3.12.12} to execute the experiments with \textit{PyTorch 2.10.0}, and \textit{NumPy 2.0.2}. The federated learning process was simulated within a single runtime using virtual clients. A random seed of 42 was fixed across the experiment where applicable, and Table~\ref{tab:parameters} presents the parameters and hyperparameters used.

\subsection{Data Collection and Pre-processing}

\subsubsection{Dataset Description}
In this study, WUSTL-IIoT-2021 dataset \cite{Zolanvari2021WUSTL}, a well-known benchmark dataset which is publicly available is used. It was developed by Washington University in St. Louis. The dataset consists of IIoT network data samples collected using a testbed \cite{8693904} that generates network traffic similar to real-world industrial systems. It is made up of 1,194,464 samples comprising 92.72\% (1,107,448) normal samples and 7.28\% (87,016) attack samples, and 49 features. The attacks include reconnaissance, command injection attacks, denial of service (DoS) attacks, and backdoors. The dataset is available in CSV format and contains 49 features comprising of both packet- and flow-level behaviors such as protocols, port numbers, packets sent and received, and flow-durations, etc.

\subsubsection{Dataset Pre-processing}
Data pre-processing represents an important step in DL/ML model training. It helps improve model efficiency, minimize errors, and stabilize the model quickly. In this study, the data samples from the dataset are cleaned and separated into numerical and categorical data types. Numerical features are normalized using \textit{z-score} through \textit{StandardScaler} feature scaling technique. Similarly, categorical features are encoded using \textit{OneHotEncoder}. The pre-processed data $\mathbf{X}_{all}$ includes both numerical and encoded categorical features and serves as input to the distributed DNAE training stage. Finally, the pre-processed data is split into train, validation, and test (80/10/10) using \textit{train\_test\_split}. The train and validation splits, which are separated into benign and  attack samples, are used during the training phase, while the test split is used for evaluation.

\subsection{Federated Learning Setup}

To reflect the distributed and heterogeneous nature of IIoT environments, we simulate FL clients as independent industrial sites. In this setup, each client holds a local dataset and participates in collaborative training without sharing raw data. We simulate a total of 10 clients where each client performs local training and hyperedges generation and communicates model updates to a central aggregator which performs weighted federated averaging (FedAvg). Moreover, the clients share the same global model architecture and optimization settings.

Furthermore, to simulate client heterogeneity, we adopt a Dirichlet-based non-IID partitioning strategy \cite{yurochkin2019bayesian} driven by protocol usage. We used a \texttt{concentration parameter ($\alpha$)} of 0.7 to induce moderate to high degree of heterogeneity without making the training process so unstable. Using lower values shows signs of potential overfitting as the interval between the training and validation reconstruction error becomes large. The WUSTL-IIoT-2021 dataset contains 8 protocol identifiers, represented numerically as \texttt{6, 17, 2054, 35020, 0, 2, 58, and 1}.

In addition to data heterogeneity, we also model partial client participation in FL setup by sampling 80\% of clients to participate in each federated round. By doing this, we aim to emulate an operational condition where by industrial sites may be temporarily offline and, altogether, mirror real IIoT deployments.

\subsection{Ablations and Baselines}

To assess the contribution of each component in the proposed EFAH-ZTM framework, we conducted a comprehensive evaluation using both ablations and a comparison with related work. First, we design the following ablation variants to isolate the impact of federated learning and hypergraph modeling:

\begin{itemize}
\item Centralized DNAE + Hypergraph + Clustering (No FL).
In this variant, we removed the FL component and trained the DNAE centrally using the aggregated dataset. However, the hypergraph construction, spectral embedding, clustering, and policy generation are applied identically to the proposed pipeline. Accordingly, we aim to quantify the effect of decentralized training on micro-segmentation outcomes in this setup.

\item Federated DNAE Only (No Hypergraph). 
We trained the DNAE in a federated manner using FedAvg in this variant, but the hypergraph construction and spectral embedding stages are removed. This means micro-segments are obtained by directly clustering the learned latent representations using the same clustering algorithms. In this ablation, we evaluated the contribution of hypergraph modeling in capturing higher-order IIoT relationships beyond latent-space similarity.

\end{itemize}

In addition to ablations, we compared the proposed EFAH-ZTM framework with two existing works that used clustering as an approach to micro-segmentation in IIoT networks. These are the works proposed by Selciya \textit{et al.} \cite{10864440} and Arifeen \textit{et al.} \cite{MurshedulArifeen}. Both studies used a centralized approach in their frameworks and used the UNSW-NB15 dataset. While we argue that IIoT network is decentralized, we compare our micro-segmentation approach with their approach.

\subsection{Performance Evaluation}
\label{subsec:perfEval}
The quality and practical utility of the discovered micro-segments are quantified rigorously using a dual set of metrics: structural quality and security efficacy.

\subsubsection{Structural Quality Metrics}
Under these metrics, we assess how well the micro-segments are defined within the spectral embedding space. Two validation metrics are used:

\begin{itemize}
    \item Silhouette Score ($\text{Sil}$): This metric measures the separation distance between the resulting micro-segments. A value close to +1 indicates that the samples under the micro-segments are tightly clustered and well-separated from neighboring micro-segments (clusters).
    \item Davies-Bouldin Index ($\text{DBI}$): This metric quantifies the average similarity ratio between each micro-segment and its most similar micro-segment. Here, lower $\text{DBI}$ value shows better micro-segmentation (clustering), indicating greater separation and denser micro-segments.
\end{itemize}

\subsubsection{Security Efficacy Metrics}
To analyze security effectiveness under labeled datasets, we report two oracle-based metrics that provide an insight into how well micro-segments separate benign and malicious behaviors. However, because these metrics are not available at runtime and rely on ground-truth labels, they are used solely for post hoc evaluation.

\begin{itemize}
    \item Micro-segment (Cluster) Purity: This metric determines the dominant traffic type (Attack or Benign) within each discovered segment. Then, purity is calculated as the proportion of samples belonging to the majority class in a segment relative to the total number of samples in that segment. This is defined in Eq.~\ref{eq:purity}.
\end{itemize}

\begin{equation}
\text{Purity}(c) = \frac{\max_{j} |\{x \in c \mid \text{class}(x) = j\}|}{|n|}
\label{eq:purity}
\end{equation}
Where:
\begin{itemize}
\item 
\begin{itemize}
    \item $|\{x \in c \mid \text{class}(x) = j\}|$ is the number of data samples in cluster $c$ belonging to class $j$ (where $j$ is typically "Attack" or "Benign" in this context).
    \item $\max_{j}$ represents the maximum number of samples found across all classes within a micro-segment $c$.
    \item $|n|$ is the total number of samples (size) in cluster $c$.
\end{itemize}
\end{itemize}

\begin{itemize}
    \item Contamination Metrics: These metrics assess the security risk through lateral-movement associated with each micro-segment. We define contamination as the proportion of the minority class within a cluster, relative to the total cluster size. This is evaluated using two approaches:

    \begin{itemize}
        \item Benign $\rightarrow$ Attack Contamination ($C_{B \rightarrow A}$): This measures the proportion of benign samples residing within a segment that is dominated by attack samples.
        \item Attack $\rightarrow$ Benign Contamination ($C_{A \rightarrow B}$): This measures the proportion of malicious attack samples residing within a segment that is dominated by benign traffic.
    \end{itemize}
\end{itemize}

For a benign-dominated cluster $c$, the Attack-to-Benign Contamination is calculated in Eq.~\ref{eq:attackTobenignCont}:

\begin{equation}
C_{A \rightarrow B}(c) = \frac{|\{x \in c \mid x \text{ is Attack}\}|}{|n|}
\label{eq:attackTobenignCont}
\end{equation}

where $|n|$ is the total number of samples in cluster $c$.

\section{Results and Discussion}
\label{sec:results}

\subsection{Experimental Overview}
In this section, we report the empirical performance of the proposed \textit{EFAH-ZTM} on the WUSTL-IIoT-2021 dataset which is based on the experimental configuration described in Section~\ref{sec:expSetup}. The evaluation is structured to reflect the key design choices in the proposed pipeline and we report the mean $\pm$ standard deviation after five runs. Similarly, we used three complementary dimensions to quantify the outcomes and these are:

\begin{itemize}
    \item We evaluate whether the embedding and clustering stages produce cohesive and well-separated segments using the Silhouette score (Sil) and Davies--Bouldin index (DBI) (Section~\ref{subsec:perfEval}). This is important for stable and interpretable policy generation.
    \item We examine and report the security efficacy under oracle evaluation using ground-truth labels to understand how well the discovered micro-segments align with benign/malicious behaviors.
    \item We evaluate the operational policy behavior is driven by the proposed \textit{operational risk score} (Section~\ref{subsec:opRisk}), and interpretability using LIME and SHAP (Section~\ref{subsec:XAIforPolJust}).
\end{itemize}

The subsequent sections present the results from the experiment in detail with corresponding discussion.

\subsection{Learning Dynamics and Computation Time}

\begin{figure}
    \centering
    \includegraphics[width=0.7\linewidth]{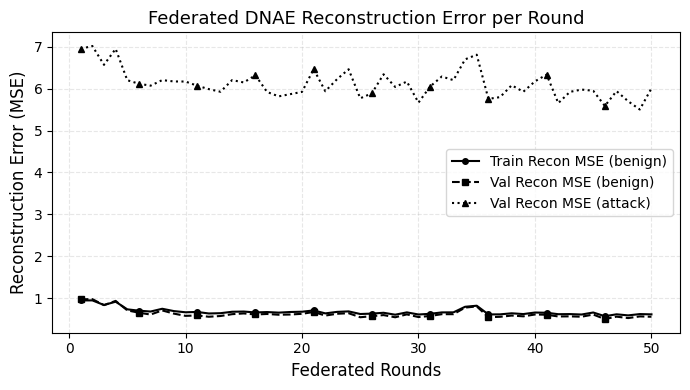}
    \caption{DNAE Reconstruction Error}
    \label{fig:recError}
\end{figure}

In the experiment, we trained a federated DNAE and its ability to reconstruct samples (reconstruction error) is a core signal used in the operational risk score.  Fig.~\ref{fig:recError} shows the reconstruction error across federated rounds for benign training traffic, benign validation traffic, and attack validation traffic. In this figure, two observations are clear. The benign curves converge to a low and stable error, which indicates that the federated optimization converges despite client heterogeneity and partial participation. However, the reconstruction error on the attack validation traffic remains consistently high and well above the benign baseline. This result supports the use of reconstruction error as an unsupervised \textit{behavioral deviation} indicator for risk-aware micro-segmentation.

Considering computation time, Table~\ref{tab:timeDuration} reports the duration in seconds of the major pipeline stages. Federated DNAE training is the dominant cost with $\approx 6908$s in the Colab simulation. Hypergraph construction and clustering, on other hand, complete lesser than that. From the table, hypergraph construction times are comparable across modes, with the manifold-based construction showing a modest overhead relative to kNN-based construction, and this could be due to diffusion/affinity computations. For clustering, HDBSCAN is faster than MiniBatch KMeans in this setup, but it also shows higher runtime variability across hypergraph modes during the experiment. The reason for this variability could be because the density-based clustering produces different numbers of clusters and neighborhood structures depending on the embedding geometry, while KMeans has a more predictable optimization path and maintains a fixed number of clusters.

\begin{table}[h]
\centering
\caption{Time Duration (Seconds)}
\label{tab:timeDuration}
\renewcommand{\arraystretch}{1.4}
\begin{tabular}{ccccc}
\hline
\multirow{2}{*}{DNAE Training} & \multicolumn{2}{c}{Hypergraph Construction} & \multicolumn{2}{c}{Cluster Formation} \\ \cline{2-5} 
 & \multicolumn{1}{c}{kNN-based} & Manifold-based & \multicolumn{1}{c}{MiniBatch KMeans} & HDBSCAN \\ \hline
$6908.46 \pm 374.56$ & \multicolumn{1}{c}{$149.28 \pm 1.60$} & $155.31 \pm 1.62$ & \multicolumn{1}{c}{$121.36 \pm 0.4689$} & $62.48 \pm 4.149$ \\ \hline
\end{tabular}
\end{table}

\subsection{Structural Quality of Micro-Segmentation}
\begin{figure}[htbp]
     \centering
     \begin{subfigure}[b]{0.49\textwidth}
         \centering
         \includegraphics[width=\linewidth]{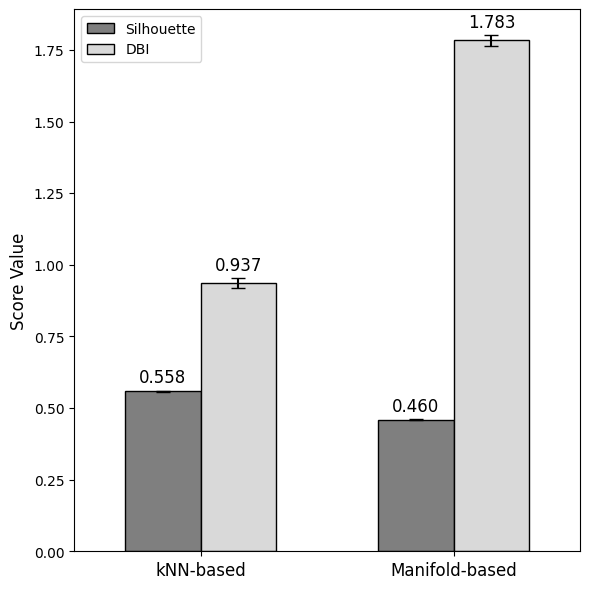}
         \caption{MiniBatch KMeans Performance}
         \label{fig:MiniBatchPerformance}
     \end{subfigure}
     \hfill
     \begin{subfigure}[b]{0.49\textwidth}
         \centering
         \includegraphics[width=\linewidth]{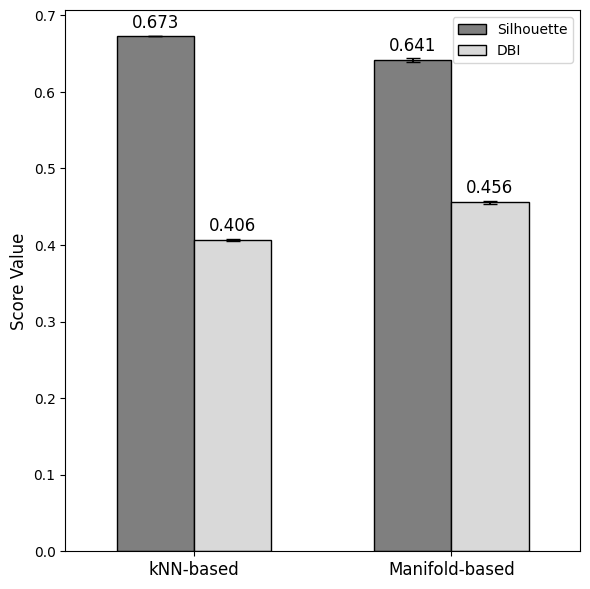}
         \caption{HDBSCAN Performance}
         \label{fig:HDBSCANPerformance}
     \end{subfigure}
     
     \caption{Structural Quality of Micro-Segments}
     \label{fig:OverallPerformance}
\end{figure}

Table~\ref{tab:clusteringPerf} reports the number of resulting micro-segments, summary statistics of cluster sizes, and clustering metrics. Fig.~\ref{fig:OverallPerformance} visualizes the Silhouette and DBI comparisons for both clusterers.

From these results, several consistent trends emerge. For example, HDBSCAN produces substantially finer-grained micro-segmentation than MiniBatch KMeans. While MiniBatch KMeans is constrained to 500 clusters by design, HDBSCAN discovers thousands of clusters (3541 for kNN hypergraph; 3185 for manifold hypergraph), with small median sizes (19--20 samples) and maxima (106--117). To interpret this in operational micro-segmentation terms, this corresponds to a more granular set of behavioral zones, which can reduce the blast radius of compromise but may increase policy-management overhead. By contrast, MiniBatch yields larger clusters with median between 191--206 micro-segments. This provides a more compact policy surface that may be preferable for administrators that need a fixed and manageable number of micro-segments.

In terms of structural quality, HDBSCAN yields stronger structural separation than KMeans in both hypergraph modes. More explicitly, the Silhouette scores for HDBSCAN are higher (0.641--0.673) than for KMeans (0.459--0.559). Moreover, the DBI is significantly lower for HDBSCAN (0.406--0.456) than for KMeans (0.937--1.783). These results align with the expected behavior of density-based clustering in heterogeneous IIoT settings in which HDBSCAN can capture variable-density and non-globular patterns that may have violated the spherical-cluster assumption of KMeans.

Comparing the hypergraph modes, kNN-based hypergraphs contribute to producing embeddings with stronger structural metrics (higher Silhouette, lower DBI) for both clusterers. This could suggest that local neighborhood hyperedges preserve compact local continuity in the latent space, which has been translated into tighter and more separable clusters in the spectral embedding. However, as shown next, structural quality alone does not fully characterize the security utility of a segmentation, and this motivates the complementary oracle security analysis.

\begin{table}[ht]
\centering
\caption{Clustering Performance: MiniBatch KMeans vs HDBSCAN}
\renewcommand{\arraystretch}{1.4}
\small
\label{tab:clusteringPerf}
\begin{tabular}{llcccc}
\toprule
\textbf{Clusterer} & \textbf{Hypergraph mode} & \textbf{Clusters} & \textbf{Size (min/med/max)} & \textbf{Sil} & \textbf{DBI} \\ 
\midrule
\multirow{2}{*}{MiniBatch KMeans} & kNN-based & 500 & 7/206/999 & $0.5585 \pm 0.0026$ & $0.9370 \pm 0.0173$ \\ \cline{2-6}
                                 & Manifold-based & 500 & 8/191/1357 & $0.4595 \pm 0.0009$ & $1.7830 \pm 0.0191$ \\ 

\midrule
\multirow{2}{*}{HDBSCAN}          & kNN-based & 3541 & 10/20/117 & $0.673 \pm 0.00$ & $0.4065 \pm 0.0009$ \\ \cline{2-6}
                                 & Manifold-based & 3185 & 10/19/106 & $0.641 \pm 0.0026$ & $0.4558 \pm 0.0022$ \\ 
\bottomrule
\end{tabular}
\end{table}

\subsection{Security Efficacy Under Oracle Evaluation}

For Zero Trust micro-segmentation, the goal is more than having good structural separation. It involves answering the question of whether the discovered segments meaningfully isolate malicious behaviors from benign operational flows, and consequently minimizing opportunities for lateral movement and limiting collateral blocking. Based on this, we report oracle security metrics (purity and contamination) using ground-truth labels (which are not available at runtime) from the dataset. However, these metrics are not used by the framework at runtime; they are reported solely as an offline evaluation to validate the security semantics of the discovered micro-segmentation.

\begin{figure}[htbp]
     \centering
     \begin{subfigure}[b]{0.49\textwidth}
         \centering
         \includegraphics[width=\linewidth]{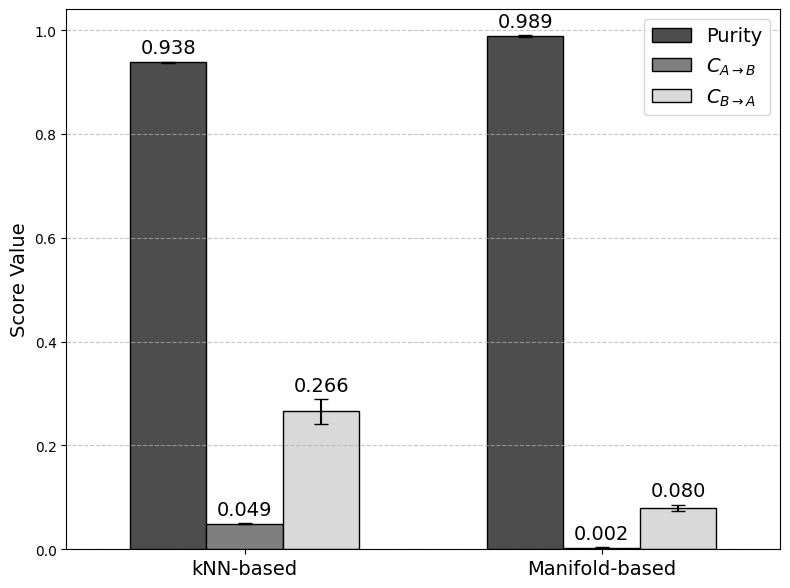}
         \caption{MiniBatch KMeans}
         \label{fig:SEuOEMiniBatch}
     \end{subfigure}
     \hfill
     \begin{subfigure}[b]{0.49\textwidth}
         \centering
         \includegraphics[width=\linewidth]{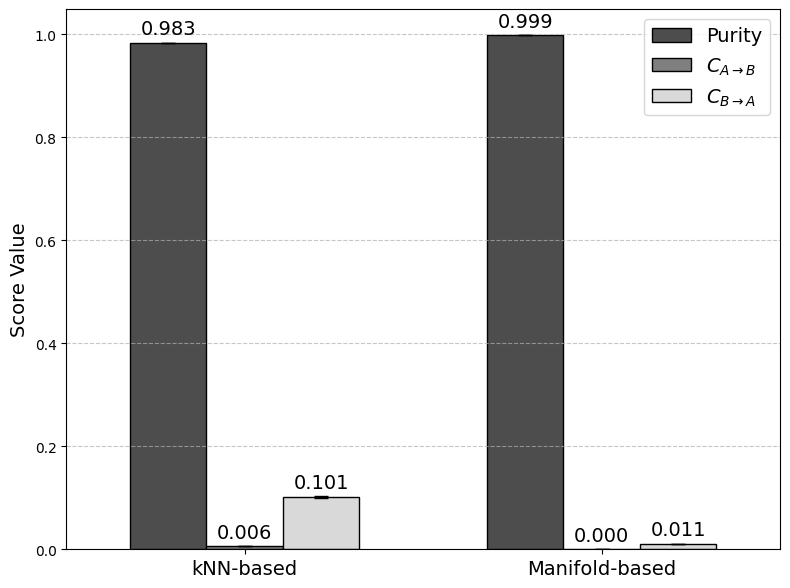}
         \caption{HDBSCAN}
         \label{fig:SEuOEHDBSCAN}
     \end{subfigure}
     
     \caption{Security Efficacy under Oracle Evaluation}
     \label{fig:SecurityEfficacyComparison}
\end{figure}

Table~\ref{tab:security_efficacy} and Fig.~\ref{fig:SecurityEfficacyComparison} show that all configurations produce high overall purity ($\geq 0.938$). This means that clusters are typically dominated by one traffic type. Similarly, the contamination metrics reveal meaningful differences in operational risk. For example, the attack $\rightarrow$ benign contamination ($C_{A \rightarrow B}$) is most critical for Zero Trust enforcement. This is because it measures the fraction of malicious samples embedded inside benign-dominated micro-segments. If the samples are mixed, it will enable an attacker to remain within an "allowed" micro-segment if the policy enforcement is purely based on membership. From the results, it is seen that the manifold-based hypergraph significantly reduces the $C_{A \rightarrow B}$ for KMeans (from $0.0493$ to $0.002$) and HDBSCAN (from $0.006$ to $0.00$). On the other hand, the benign $\rightarrow$ attack contamination ($C_{B \rightarrow A}$) captures the degree by which benign operational flows are absorbed into attack-dominated segments. The effect of this form of contamination is that it can result in unnecessary blocking and operational disruption if attack-dominated segments are isolated. Similarly, the manifold-based hypergraph also reduces this effect down to $0.0110$ for HDBSCAN.

Overall, it can be said that the HDBSCAN with manifold-based hypergraph construction provides the strongest oracle-aligned security semantics. It achieves near-perfect purity ($0.9990$) and had both contamination measures close to zero. Another important insight from the results is that the configuration with the best structural metrics may not always be the best under oracle security metrics. For example, while kNN-based hypergraphs generally yield higher Silhouette scores, manifold-based hypergraphs show lower contamination. This could suggest that diffusion-based hyperedges better preserve global separation between behavioral regimes (benign vs. attack) even when local cluster compactness is reduced.

\begin{table}[ht]
\centering
\caption{Security Efficacy under Oracle Evaluation}
\renewcommand{\arraystretch}{1.4}
\small
\label{tab:security_efficacy}
\begin{tabular}{llccc}
\hline
\multicolumn{1}{c}{\textbf{Clusterer}} & \multicolumn{1}{c}{\textbf{Hypergraph mode}} & \textbf{Purity} & \textbf{$C_{A \rightarrow B}$} & \textbf{$C_{B \rightarrow A}$} \\ \hline
\multirow{2}{*}{MiniBatch KMeans} & kNN-based & $0.9380 \pm 0.0014$ & $0.0493 \pm 0.0011$ & $0.2655 \pm 0.0243$ \\ \cline{2-5}
& Manifold-based & $0.9888 \pm 0.0022$ & $0.002 \pm 0.0017$ & $0.0803 \pm 0.0056$ \\ \hline
\multirow{2}{*}{HDBSCAN} & kNN-based & $0.9833 \pm 0.0004$ & $0.006 \pm 0.00$ & $0.1010 \pm 0.0017$ \\ \cline{2-5}
& Manifold-based & $0.9990 \pm 0.00$ & $0.00 \pm 0.00$ & $0.0110 \pm 0.00$ \\ \hline
\end{tabular}
\end{table}

\subsection{Operational Risk--Based Policy behavior}
In the proposed EFAH-ZTM framework, micro-segmentation policies generation is driven by the operational risk score (Section~\ref{subsec:opRisk}). This aggregates (i) reconstruction-error deviation from the federated DNAE and (ii) structural outlierness with respect to the discovered micro-segments. At runtime, these instance-level scores are aggregated to cluster-level risk values $R(c)$, and which guides the policy decision for intra-segment communication.

\begin{figure}
    \centering
    \includegraphics[width=1.0\linewidth]{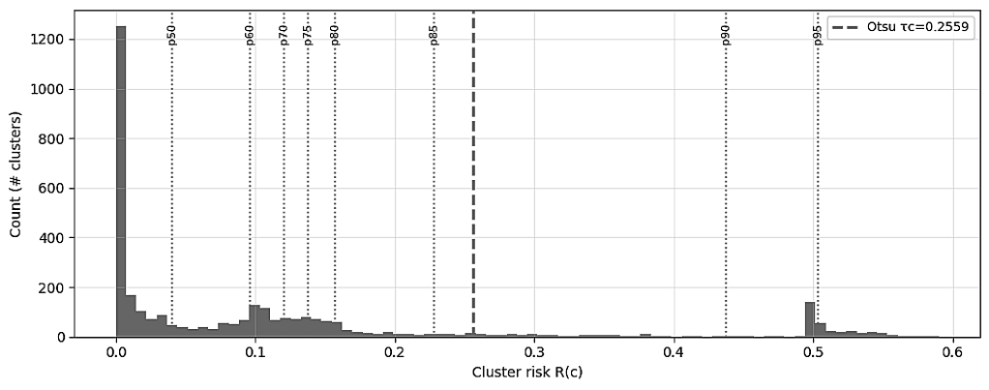}
    \caption{Operational Risk Score Distribution Over Micro-segments}
    \label{fig:Otsu}
\end{figure}

Fig.~\ref{fig:Otsu} shows the empirical distribution of $R(c)$ over the discovered micro-segments using the HDBSCAN clusterer. The figure highlights several decision candidate thresholds $\tau_c$ derived using Otsu's method and percentiles. From the result shown, it means that if we are to select, for example, the threshold from the Otsu's method, then the intra-segment communication for more than 85\% of micro-segments with risk scores $\leq$ 0.2559 will be allowed, separating a large mass of low-risk micro-segments from a small high-risk set. In effect, this behavior is consistent with the finer granularity and tighter structural separation achieved by HDBSCAN. 

This demonstrates the intended policy design: risk-based intra-segment allow rules preserve operational continuity for low-risk behavioral groups, while segment isolation is activated only for micro-segments exhibiting high deviation. At the same time, the default-deny inter-segment rule provides an explicit lateral-movement barrier which is independent of the chosen threshold. From the experiment, the average operational risk threshold (Otsu’s threshold) for MiniBatch KMeans and HDBSCAN are  $0.3936 \pm 0.0192$ ($> p75$) and $0.2356 \pm 0.0204$  ($> p85$), respectively.

\subsection{Explainability and Policy Interpretability}
A key component and objective of the proposed EFAH-ZTM framework is, not only generate micro-segmentation policies but, to provide a feature-level explanation for security operators. To achieve this, we used a surrogate classifier (kNN model) trained on the spectral embeddings which are obtained from the test split of the dataset used to approximate the clustering boundary, and this enables post hoc explanation using LIME and SHAP techniques.

The surrogate classifier achieves strong predictive fidelity with respect to cluster assignments:
\[
\text{accuracy} = 0.9927 \pm 0.0027, \quad
\text{macro F1} = 0.9874 \pm 0.0051, \quad
\text{micro F1} = 0.9927 \pm 0.0027.
\]
Accordingly, this level of performance indicates that the surrogate provides a reliable interface for explanation. That is, the derived feature attributions are grounded in a decision boundary that closely matches the produced micro-segmentation.

We further evaluate the consistency of the explanations through five runs. As expected, SHAP explanations with fixed model and background are essentially deterministic, and this gives a stability of $1.0 \pm 0.00$. On the other hand, LIME exhibits little variability due to its perturbation-based sampling procedure, but still remains highly stable overall ($0.9659 \pm 0.0341$). Examining the stability and achieving high score is important because in operational settings, it reduces analyst fatigue and supports repeatable audits. In other words, the same policy decision should yield similar justification factors under repeated explanation runs.

\begin{figure*}[h!]
    \centering
    \begin{subfigure}[b]{0.7\textwidth}
        \centering
        \includegraphics[width=\textwidth]{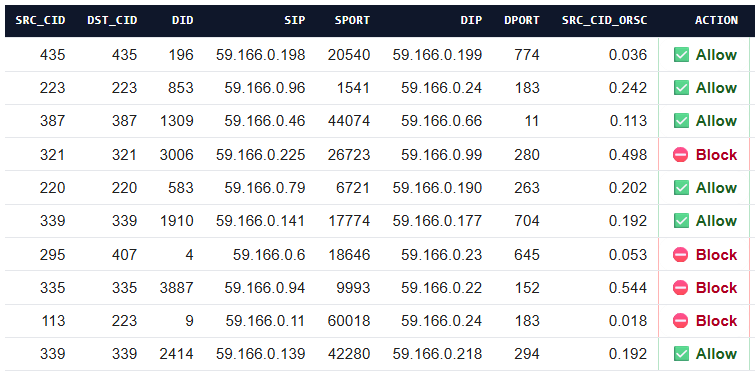}
        \caption{Left Half Portion of the Policy Table}
        \label{fig:XAI_Results_A}
    \end{subfigure}
    
    \hfill 
    \begin{subfigure}[b]{0.95\textwidth}
        \centering
        \includegraphics[width=\textwidth]{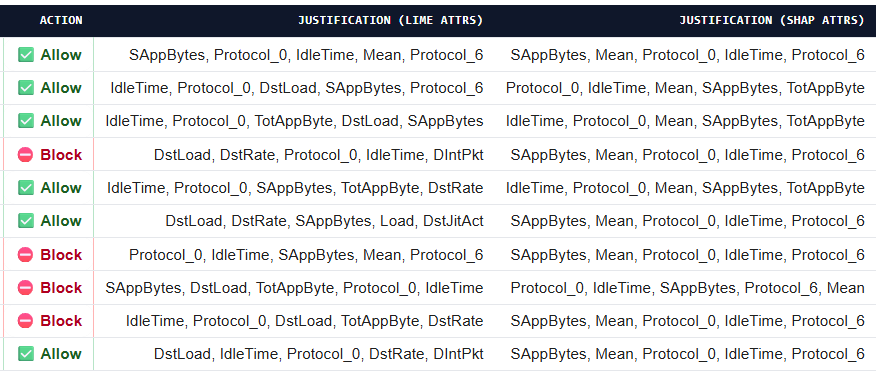}
        \caption{Right Half Portion of the Policy Table}
        \label{fig:XAI_Results_B}
    \end{subfigure}
    
    \caption{XAI Policy Table}
    \label{fig:XAI_Results}
\end{figure*}

Fig.~\ref{fig:XAI_Results} illustrates the structure of the final XAI-augmented policy table. The table combines several important components: (i) segment identifiers/attributes (Source Cluster ID (SRC\_CID), Destination Cluster ID (DST\_CID), and Source Cluster Operational Risk Score (SRC\_CID\_ORSc)), (ii) device identifiers (Device ID (DID), Source IP (SIP), and Destination IP (DIP)), (iii) flow attributes (Source Port (SPort) and Destination Port (DPort)), (iv) an Allow/Block decision produced by the policy logic, and (v) a compact explanation in the form of top contributing features from LIME and SHAP. The DID represents the sample number from the dataset. 

In practice, the justifications obtained from the LIME and SHAP techniques can serve two complementary roles. They can help operators to understand which traffic characteristics are driving the micro-segmentation. Also, they can help in verifying that segmentation aligns with expected operational semantics, for example, protocol usage, traffic-volume patterns, or timing/idle behavior and detect mis-segmentation or drift.

\subsection{Ablation and Baseline Analysis}
In this section, we evaluate two key components of our proposed EFAH-ZTM framework through the following ablation studies: (i) centralized training (No FL), and (ii) removal of hypergraph modeling (No Hypergraph). The goal here is not to claim universal superiority of one setting, but to clarify what each component contributes in terms of structure, security semantics, and operational feasibility.

\subsubsection{Centralized DNAE + Hypergraph + Clustering (No FL)}
We trained the DNAE model in a centralized manner. I.e., instead of distributing the dataset into a number of clients and aggregating their updates globally, the whole dataset is considered as one big chunk and a single model is trained. Effectively, this makes the training completely under homogeneous dataset. Under this setting, the DNAE model attains lower reconstruction error for benign traffic (train\_ben\_mse=0.2898 and val\_ben\_mse=0.28684) compared to the FL DNAE while remaining highly sensitive to attacks (val\_att\_mse=5.8221). It also further highlights that reconstruction-based deviation remains a strong unsupervised risk signal even without FL.

Structurally, Table~\ref{tab:clusteringPerfCentral} shows that centralized training can improve KMeans separation under the kNN hypergraph, where Silhouette increases from $0.5585$ to $0.7025$. However, HDBSCAN remains broadly similar across the centralized and federated training. In terms of oracle security efficacy as shown in Table~\ref{tab:security_efficacyCentral}, while the centralized training provides mixed contamination outcome, the federated setting remains competitive. And in several HDBSCAN configurations, it yields lower contamination than the centralized training. This in effect suggests that decentralization does not inherently degrade micro-segmentation quality, rather it provides clear deployment advantages in IIoT environments where data centralization is constrained by factors such as privacy, governance, and connectivity considerations.

Furthermore, the operational risk score distribution across micro-segments under the centralized training is largely similar to the federated setting with only a major difference in the percentile coverage using MiniBatch KMeans clustering method. It increases from $>p75$ in the FL to $>p85$ which means more intra-segment communication will be allowed. The average Otsu’s threshold for the two clustering techniques, MiniBatch KMeans and HDBSCAN, are $0.3927 \pm 0.0055$ ($> p85$) and $0.2291 \pm 0.0017$  ($> p80$), respectively.

\begin{table}[ht]
\centering
\caption{Clustering Performance (Centralized Training): MiniBatch KMeans vs HDBSCAN}
\renewcommand{\arraystretch}{1.4}
\small
\label{tab:clusteringPerfCentral}
\begin{tabular}{llcccc}
\toprule
\textbf{Clusterer} & \textbf{Hypergraph mode} & \textbf{Clusters} & \textbf{Size (min/med/max)} & \textbf{Sil} & \textbf{DBI} \\ 
\midrule
\multirow{2}{*}{MiniBatch KMeans} & kNN-based & 500 & 3/209/854 & $0.7025 \pm 0.0188$ & $0.9270 \pm 0.0180$ \\ \cline{2-6}
                                 & Manifold-based & 500 & 2/209/1024 & $0.4837 \pm 0.0009$ & $1.4253 \pm 0.0179$ \\ 

\midrule
\multirow{2}{*}{HDBSCAN}          & kNN-based & 3626 & 10/19/112 & $0.674 \pm 0.00$ & $0.385 \pm 0.00$ \\ \cline{2-6}
                                 & Manifold-based & 2905 & 10/20/121 & $0.6280 \pm 0.0020$ & $0.4410 \pm 0.0007$ \\ 
\bottomrule
\end{tabular}
\end{table}

\begin{table}[ht]
\centering
\caption{Security Efficacy under Oracle Evaluation (Centralized Training)}
\renewcommand{\arraystretch}{1.4}
\small
\label{tab:security_efficacyCentral}
\begin{tabular}{llccc}
\hline
\multicolumn{1}{c}{\textbf{Clusterer}} & \multicolumn{1}{c}{\textbf{Hypergraph mode}} & \textbf{Purity} & \textbf{$C_{A \rightarrow B}$} & \textbf{$C_{B \rightarrow A}$} \\ \hline
\multirow{2}{*}{MiniBatch KMeans} & kNN-based & $0.9450 \pm 0.00$ & $0.0688 \pm 0.0018$ & $0.00 \pm 0.00$ \\ \cline{2-5}
& Manifold-based & $0.985 \pm 0.00$ & $0.006 \pm 0.00$ & $0.0198 \pm 0.00$ \\ \hline
\multirow{2}{*}{HDBSCAN} & kNN-based & $0.9640 \pm 0.00$ & $0.011 \pm 0.00$ & $0.174 \pm 0.00$ \\ \cline{2-5}
& Manifold-based & $0.9960 \pm 0.00$ & $0.001 \pm 0.00$ & $0.035 \pm 0.00$ \\ \hline
\end{tabular}
\end{table}

\subsubsection{Federated DNAE Only (No Hypergraph)}
The second component of the ablation study is removing the hypergraph stage from the pipeline. This effectively isolates the effect of higher-order relationship modeling and spectral embedding. As shown in Table~\ref{tab:clustering_performanceNoHypergraph}, the structural quality degrades significantly without hypergraphs, particularly for MiniBatch KMeans (Silhouette drops from $0.5585$ to $0.3297$). For HDBSCAN, both Silhouette decreases and DBI increases compared to the hypergraph-based variants. This indicates weaker separation in the direct latent space than in the hypergraph spectral space.

By contrast, the oracle security metrics remain high without hypergraphs as shown in Table~\ref{tab:security_efficacyNoHypergraph}, reflecting that the federated latent space already captures strong benign/attack separability. However, the absence of hypergraph modeling consequently changes the granularity and topology of segmentation. It is observed that HDBSCAN produces far fewer clusters (409) and much larger cluster sizes (median 56). The effect of this is that it can enlarge the operational attack surface by placing more endpoints/flows within the same micro-segment. Moreover, the Otsu thresholds shift (KMeans: $0.5014$ at $<p60$), which suggests a less sharply bi-modal risk separation that may lead to broader blocking under strict enforcement. This ablation demonstrates that the hypergraph shapes the structure, granularity, and risk stratification of micro-segmentation in ways that align with Zero Trust objectives of, for example, limiting  attack surface.

\begin{table}[ht]
    \centering
    \caption{Clustering Performance (No Hypergraph): MiniBatch KMeans vs HDBSCAN}
    \label{tab:clustering_performanceNoHypergraph}
    \begin{tabular}{lcccc}
        \toprule
        \textbf{Clusterer} & \textbf{Number of Clusters} & \textbf{Cluster Size (min/med/max)} & \textbf{Sil} & \textbf{DBI} \\ 
        \midrule
        MiniBatch KMeans & 500 & 10/173/869 & $0.3297 \pm 0.0009$ & $0.997 \pm 0.00$ \\
        HDBSCAN          & 409 & 25/56/427   & $0.5440 \pm 0.00$   & $0.577 \pm 0.00$ \\
        \bottomrule
    \end{tabular}
\end{table}

\begin{table}[ht]
    \centering
    \caption{Security Efficacy under Oracle Evaluation (No Hypergraph)}
    \label{tab:security_efficacyNoHypergraph}
    \begin{tabular}{lccc}
        \toprule
        \textbf{Clusterer} & \textbf{Purity} & \textbf{$C_{A \rightarrow B}$} & \textbf{$C_{B \rightarrow A}$} \\ 
        \midrule
        MiniBatch KMeans & $0.9970 \pm 0.00$ & $0.007 \pm 0.00$ & $0.022 \pm 0.00$ \\
        HDBSCAN          & $0.9980 \pm 0.00$ & $0.001 \pm 0.00$ & $0.002 \pm 0.00$ \\
        \bottomrule
    \end{tabular}
\end{table}


\subsubsection{Baseline Comparison with Prior Work}

In this section, we compare the proposed EFAH-ZTM framework with two relevant micro-segmentation baselines: the OPTICS+Decision Tree pipeline proposed by Arifeen \textit{et al.}~\cite{MurshedulArifeen} and the DNAE + kNN-hypergraph clustering approach proposed by Selciya \textit{et al.}~\cite{10864440}. These studies represent two common directions in prior work which are density-based clustering with supervised enforcement, and representation learning enhanced by hypergraph structure. 

In \cite{MurshedulArifeen}, Arifeen \textit{et al.} first used OPTICS to form micro-segments and later applied a Decision Tree classifier to label traffic as normal or malicious for policy generation. Based on their experiments on UNSW-NB15 and IoTID20 datasets, it shows that the method can support the prevention of lateral-movement through restricting redundant communication links within each segment. However, OPTICS did not scale well in their setting and, therefore, the clustering (178 clusters with UNSW-NB15 and 295 clusters with IoTID20) was performed on only 1,000 randomly selected samples. This scalability issue becomes a limiting factor in practical large IIoT traffic volumes. In contrast to this supervised and label-dependent approach, EFAH-ZTM uses an unsupervised operational risk score that is based on behavioral deviation (reconstruction error) and structural outlierness in the segmentation space. This difference is important in IIoT contexts where labels are usually unavailable or incomplete, and where the threat landscape could include previously unseen behaviors.

Similarly, in \cite{10864440}, Selciya \textit{et al.} used DNAE and kNN-based hypergraph clustering to better capture higher-order traffic relationships. The results from their experiment on a randomly selected 1,000-sample subsets of UNSW-NB15 show that adding hypergraph improves clustering quality over PCA or latent-space similarity alone. EFAH-ZTM builds on this direction by (i) extending hypergraph construction to both kNN-based and manifold/diffusion-inspired modes. This enables the micro-segmentation process to capture non-linear connectivity along the underlying behavioral manifold in addition to local neighbor similarity; (ii) introducing federated learning to avoid raw-data centralization suitable for distributed industrial deployments; (iii) proposing risk-aware Zero Trust policy logic for intra-segment communication which addresses micro-segments with potential malicious devices; and (iv) adding LIME/SHAP XAI techniques for feature-level explanations of generated micro-segmentation policies.

\subsection{Limitations and Future Work}
The results presented and discussed indicate that the proposed EFAH-ZTM framework can produce micro-segmentation policies that are high-quality, security-relevant, and operationally interpretable  and consistent with Zero Trust requirements. However, there are notable limitations that should be acknowledged that define potential directions for future research. These include:

\begin{itemize}
    \item The empirical evaluation was conducted on the WUSTL-IIoT-2021 dataset under a controlled experimental setting. While this benchmark is suitable for reproducible analysis, it may not fully represent the diversity of real industrial deployments where device populations, protocols, attack behaviors, and operational workflows vary significantly in different sectors. Accordingly,  future work should therefore validate EFAH-ZTM on additional datasets and on geographically distributed industrial testbeds or live pilot environments.
    \item The cluster-level risk is computed through aggregation over instance risks. While this improves stability, rare malicious behaviors inside otherwise benign clusters can be diluted under mean aggregation. Therefore, robust alternatives are important extensions when environments demand aggressive containment of sparse attacks.
    \item The current threat model excludes adversarial manipulation of the FL process. Therefore, incorporating secure aggregation and robust FL defenses would further strengthen deployment in hostile settings.
    \item While the explainability module in the framework improves transparency, future work could incorporate counterfactual and global explanation methods, and also evaluate the usability of the generated explanations with human analysts in realistic security operation workflows.
    \item The framework adopts a conservative Zero Trust policy where inter-segment communication is denied by default, and intra-segment communication is permitted only when segment-level operational risk is below a selected threshold. This could be too  restrictive for industrial processes that require tightly coordinated cross-segment communication. Hence, an important future research direction could incorporate context-aware policy exceptions, and process-aware dependency modeling.
\end{itemize}

\section{Conclusion}
\label{sec:conclusion}

The proposed EFAH-ZTM combines federated behavioral representation learning, hypergraph-based modeling of complex higher-order relationships, risk-aware policy generation, and explainable AI within the micro-segmentation workflow. Our experimental results show that this combination can produce micro-segments that are not only coherent in structure, but also relevant from a security perspective, as well as interpretable for operational use. Moreover, the ablation study conducted indicates that the federated setting remains competitive with centralized training, while the hypergraph modeling included in the pipeline improves the micro-segmentation structure, and risk categorization. Overall, these findings suggest that micro-segmentation in IIoT can be made more adaptive and more explainable without giving up the distributed learning setting required in many industrial deployments. In future research work, the following are important directions for improvement. These include improving robustness in the federated training, incorporating temporal behavioral changes, validating on additional datasets and on geographically distributed industrial testbeds, and refining the policy generation with richer operational context to better support real-world industrial environments.

\bibliographystyle{elsarticle-harv} 
\bibliography{references}

\end{document}